\documentclass[]{rQUF2e}
\usepackage{multirow,setspace,amssymb,amsmath,graphicx,color,rotating,subfigure,url,lineno,natbib}
\usepackage{textcomp}
\usepackage{array}
\newcommand{\PreserveBackslash}[1]{\let\temp=\\#1\let\\=\temp}
\newcolumntype{C}[1]{>{\PreserveBackslash\centering}p{#1}}
\newcolumntype{R}[1]{>{\PreserveBackslash\raggedleft}p{#1}}
\newcolumntype{L}[1]{>{\PreserveBackslash\raggedright}p{#1}}

\bibpunct{(}{)}{,}{a}{}{,}
\begin{document}
\doi{10.1080/1469768YYxxxxxxxx}
 \issn{1469-7696} \issnp{1469-7688} \jvol{00} \jnum{00} \jyear{2008} \jmonth{July}

\markboth{Fei Ren, Ya-Nan Lu, Sai-Ping Li, Xiong-Fei Jiang, Li-Xin Zhong and Tian Qiu}{Dynamic portfolio strategy using clustering approach}

\title{Dynamic portfolio strategy using clustering approach}

\author{Fei Ren$^{\ast}$${\dag}$\thanks{$^\ast$Corresponding author. Email: fren@ecust.edu.cn}
\vspace{12pt} Ya-Nan Lu$\dag$
\vspace{12pt} Sai-Ping Li$\ddag$
\vspace{12pt} Xiong-Fei Jiang$\S$
\vspace{12pt} Li-Xin Zhong$\sharp$
\vspace{12pt} and Tian Qiu$\natural$
\\\vspace{12pt}  \normalfont{$\dag$School of Business, East China University of Science and Technology, Shanghai 200237, China\\$\ddag$Institute of Physics, Academia Sinica, Taipei 115 Taiwan\\$\S$College of Information Engineering, Ningbo Dahongying University, Ningbo 315175, China\\$\sharp$School of Finance, Zhejiang University of Finance and Economics, Hangzhou 310018, China\\$\natural$School of Information Engineering, Nanchang Hangkong University, Nanchang 330063, China}
\\\vspace{12pt} }

\maketitle
\begin{abstract}
The problem of portfolio optimization is one of the most important issues in asset management. This paper proposes a new dynamic portfolio strategy based on the time-varying structures of MST networks in Chinese stock markets, where the market condition is further considered when using the optimal portfolios for investment. A portfolio strategy comprises two stages: selecting the portfolios by choosing central and peripheral stocks in the selection horizon using five topological parameters, i.e., degree, betweenness centrality, distance on degree criterion, distance on correlation criterion and distance on distance criterion, then using the portfolios for investment in the investment horizon. The optimal portfolio is chosen by comparing central and peripheral portfolios under different combinations of market conditions in the selection and investment horizons. Market conditions in our paper are identified by the ratios of the number of trading days with rising index or the sum of the amplitudes of the trading days with rising index to the total number of trading days.
We find that central portfolios outperform peripheral portfolios when the market is under a drawup condition, or when the market is stable or drawup in the selection horizon and is under a stable condition in the investment horizon. We also find that the peripheral portfolios gain more than central portfolios when the market is stable in the selection horizon and is drawdown in the investment horizon. Empirical tests are carried out based on the optimal portfolio strategy. Among all the possible optimal portfolio strategy based on different parameters to select portfolios and different criteria to identify market conditions, $65\%$ of our optimal portfolio strategies outperform the random strategy for the Shanghai A-Share market and the proportion is $70\%$ for the Shenzhen A-Share market.

\begin{keywords} Econophysics; Stock network; Portfolio

\end{keywords}
\end{abstract}
\vspace{12pt}

\section{Introduction}
\label{sec:Intro}

Portfolios is one of the hottest issues in the financial areas. It primarily concerns with the best combination of securities for specific profits that investors need. The fundamental theory about portfolio optimization can be chased back to Markowitz framework \citep{Markowitz-1952-JF}, which select the allocation of investors' investment based on a mean-variance analysis.

Much effort has been put to solve and expand the Markowitz model. In the original Markowitz model, the risk is measured by the standard deviation or variance. Several other risk measures have been later considered, creating a family of mean-risk models, e.g., mean-absolute deviation\citep{Konno-Yamazaki-1991-MS,Feinstein-Thapa-1993-MS,Simaan-1997-MS}, semi-variance \citep{Markowitz-1959,Choobineh-Branting-1986-EJOR}, Gini's mean difference \citep{Yitzhaki-1982-AER}, CVaR \citep{Konno-Waki-Yuuki-2002-APFM} and worst-case CVaR \citep{Zhu-Fukushima-2009-OR}. To deal with the noise and error caused by historic sample data in estimation of the mean, variance and also the covariance among returns, many improved estimation methods has been used, such as Bayesian estimators \citep{Frost-Savarino-1986-JFQA,Jorion-1986-JFQA}, shrinkage estimators \citep{Ledoit-Wolf-2003-JEF,Ledoit-Wolf-2004-JMA}, and fuzzy set method \citep{Tanaka-Guo-1999-EJOR,ArenasParra-BilbaoTerol-RodriguezUria-2001-EJOR}.

Although the Markowitz model seems ideal in theory, many constraints involved indeed influence the portfolios in practical investments. \citet{Black-Litterman-1992-FAJ} have found that short-sale constraints often cause no investment in many stocks, and \citet{Jagannathan-Ma-2003} have found that constraining portfolio weights to be nonnegative can reduce the risk in estimated optimal portfolios even when the constraints are wrong. To solve this problem, \citet{DeMiguel-Garlappi-Nogales-Uppal-2009-MS} proposed a generalized approach to optimize portfolios under general norm-constrained portfolio.

Many studies have concentrated on the out-of-sample performance of minimum variance portfolios of Markowitz model. Comparison among different weight distributions have found that minimum variance portfolios are at least the same good as the others even when taking returns into consideration \citep{Jorion-1986-JFQA,Jorion-1991-JBF}, and equal weight portfolio($1/N$ rule) usually turns out to be better than those more complicated one \citep{DeMiguel-Garlappi-Uppal-2009-RFS}. By comparing the efficiency of the equal weight portfolio and the minimum variance portfolios, \citet{Duchin-Levy-2009-JPM} have found that $1/N$ rule outperforms the mean-variance rule for individual small portfolios out of sample, but not for large portfolios. In addition, multi-objective portfolios\citep{Zopounidis-Doumpos-2013-Top,Lee-Chesser-1980-JPM} and multiperiod investment \citep{Hakansson-1971-JF,Li-Ng-2000-MF} have also been proposed and carefully studied.

A great deal of works have subsequently contributed to the study of portfolios by using a variety of alternative methods, e.g., neural networks \citep{Fernandez-Gomez-2007-COR,Ko-Lin-2008-ESA,Nazemi-Abbasi-Omidi-2015-AI}, genetic algorithms \citep{Chen-Mabu-Hirasawa-2010-COR,Chen-Hirasawa-2011-IEEEJtee}, simulated annealing \citep{Crama-Schyns-2003-EJOR}, Random Matrix Theory (RMT) filtering \citep{Daly-Crane-Ruskin-2008-PA,Dai-Xie-Jiang-Jiang-Zhou-2016-EmpE}, and hierarchical clustering \citep{Onnela-Chakraborti-Kaski-Kertesz-Kanto-2003-PRE,Pozzi-DiMatteo-Aste-2013-SR,Nanda-Mahanty-Tiwari-2010-ESA,Liao-Chou-2013-ESA}. Among these methods, hierarchical clustering is one of the most efficient methods for the selection of a basket of stocks for optimal portfolios. In fact, the selection of a set of stocks is pre-requisite for the Markowitz theory \citep{Pai-Michel-2009-IEEEtec,Nanda-Panda-2014-SEC}, since it is dedicated to the investment proportion of a limited number of selected stocks.

By using the hierarchical clustering method, the correlations between shares are revealed by the topological structure of the constructed stock network \citep{Mantegna-1999-EPJB,Onnela-Kaski-Kertesz-2004-EPJB,Tumminello-Aste-DiMatteo-Mantegna-2005-PNAS,
Tumminello-DiMatteo-Aste-Mantegna-2007-EPJB,Brida-Risso-2007-IJMPC,Garas-Argyrakis-2009-EPL,
Aste-Shaw-Matteo-2010-NJP,Tumminello-Lillo-Mantegna-2010-JEBO,Kwapien-Drozdz-2012-PR,Yang-Zhu-Li-Chen-Deng-2015-PA}, and can be further applied to portfolio optimization. A minimum spanning tree (MST) \citep{West-1996} description of the correlations between stocks has shown that the stocks included in the minimum risk portfolio (the optimal Markowitz portfolio) tend to lie on the outskirts of the asset network \citep{Onnela-Chakraborti-Kaski-Kertesz-Kanto-2003-PRE,Onnela-Chakraborti-Kaski-Kertesz-2002-EPJB}. By extracting the dependency structure of financial equities using both MST and planar maximally filtered graphs (PMFG) methods, it has been found that portfolios set up from a selection of peripheral stocks have lower risk and better returns than portfolios set up from a selection of central stocks, in which the centrality/peripherality is measured by indices like degree, betweenness centrality, eccentricity, closeness and eigenvector centrality \citep{Pozzi-DiMatteo-Aste-2013-SR}. A K-means clustering algorithm and its extension C-means clustering algorithm are applied to the classification of stocks, and the stocks selected from these classified groups are used for building portfolios, which perform better than the benchmark index \citep{Nanda-Mahanty-Tiwari-2010-ESA}. Similar works have been done in India, Taiwan and China stock markets using the same K-means cluster analysis \citep{Liao-Chou-2013-ESA,Pai-Michel-2009-IEEEtec}. Furthermore, clusters or communities detected based on network graphics can also provide useful information for correlations among stocks \citep{Tumminello-DiMatteo-Aste-Mantegna-2007-EPJB,Aste-Shaw-Matteo-2010-NJP,Song-Tumminello-Zhou-Mantegna-2011-PRE,Jiang-Chen-Zheng-2014-SR}, which has also been applied in the stock selection of portfolios \citep{Boginski-Butenko-Shirokikh-Trukhanov-Lafuente-2014-AOR,Choudhury-Ghosh-Bhattacharya-Fernandes-Tiwari-2014-Neurocomputing,Ross-2014-PRE}. This type of method is analogous to the clustering algorithm for their similar way in stock selection from clusters or communities partitioned by specific approaches.

In recent studies, more and more evidences show that the topological structures of stock networks are evolving over time and changes markedly during financial crises \citep{Aste-Shaw-Matteo-2010-NJP,Song-Tumminello-Zhou-Mantegna-2011-PRE,Fenn-Porter-Williams-McDonald-Johnson-Jones-2011-PRE,
Drozdz-Grummer-Gorski-Ruf-Speth-2000-pa,Podobnik-Wang-Horvatic-Grosse-Stanley-2010-EPL,Kenett-Tumminello-Madi-GurGershgoren-Mantegna-BenJacob-2010-PLoS1,
Kenett-Raddant-Lux-BenJacob-2012-PLoS,Ren-Zhou-2014-PLoS,Jiang-Chen-Zheng-2014-SR}. Therefore, a fixed set of stocks is not a wise choice for the portfolio selection under different market conditions. One possible way of solving this problem is to identify the stock clusters based on the network graphics in different time periods (moving windows) and then do the portfolio selection from the identified clusters in each period \citep{Pozzi-DiMatteo-Aste-2013-SR}. An alternative way is to use the dynamic conditional correlations (DCC) method to estimate time-varying correlations among stock returns based on the Markowitz framework \citep{Case-Yang-Yildirim-2012-JREFE,KotkatvuoriOrnberg-Nikkinen-Aijo-2013-IRFA,MirallesMarcelo-MirallesQuiros-MirallesQuiros-2015-NAJEF}. A new estimator called detrended cross-correlation coefficients (DCCA) is also used to describe the correlations between nonlinear dynamic series \citep{Sun-Liu-2016-PA,Podobnik-Stanley-2008-PRL,Zhou-2008-PRE,Jiang-Zhou-2011-PRE}. Other works refer to evolutionary algorithms include \citep{Pai-Michel-2009-IEEEtec,Suganya-Pai-2012-IJSystSci}.

The main motivation of this paper is to propose a new dynamic portfolio strategy based on the time-varying structures of the financial filtered networks in Chinese stock markets. A moving window with size $\delta t$ is used to study the variance of stock networks over time $t$. We choose the MST method to filter out the network graph in each window for its validity and simplicity, which is generated by connecting the nodes with most important correlations. The portfolio selection is determined by the network structure in the previous window (selection horizon), which is picked from a selection of peripheral stocks, most diverse corresponding to the Markowitz portfolio with minimum variance \citep{Pozzi-DiMatteo-Aste-2013-SR}, and central stocks, highly correlated and synchronous in price movements. The selected portfolios are subsequently used for investment in the following investment horizon.

The underlying market conditions are further considered in our dynamic portfolio strategy, which composes the investment strategy together with the portfolio selection. A recent study has verified that accurate price and volatility predictions can be used as a basis for the particular trading strategy adopted for portfolio \citep{Choudhury-Ghosh-Bhattacharya-Fernandes-Tiwari-2014-Neurocomputing}. In our work, we suppose that the optimal portfolio can change under different market conditions, and portfolio investments are implemented based on both historical price change and price prediction in future. For simplicity, three market conditions: drawup and drawdown trends of the daily price and a relatively stable status in between are identified respectively in the selection and investment horizons. A variety of selected portfolios are compared under different combinations of market conditions in the two horizons, and the optimal portfolio with the largest profit is found out in each market condition. To further testify the efficiency of our dynamic portfolio strategy, we perform a training using the first half of sample data, and use the optimal portfolio obtained from training to do investments using the remaining half data. Our results show that the optimal portfolio outperform the benchmark stock index on average.

The paper is organized as follows. In Section 2, we introduce our database and their summary statistics. Sections 3 provides the methods used in our study, including network construction, portfolio selection, and determination of investment horizon and market condition. The results of the optimal portfolios and their relevant performances are presented in Section 4. Section 5 summarizes our findings.

\section{Data}

Our daily data include $181\times 2$ stocks listed on Shanghai and Shenzhen A-Share markets, which have the largest volumes in two major stock exchanges in mainland China, over the period of 15 years from January 1, 2000 to December 31, 2014. To ensure the continuity and integrity of the data, the stocks selected in our study are most actively traded stocks throughout the sample period. For this purpose, we filter out those stocks which were once suspended from the market for more than 46 trading days, about 1\% of a total of 3,627 trading days. This filtering yields the sample data including 181 A-Share stocks for each market.

\begin{table}
\begin{center}
\begin{minipage}[c]{1\linewidth}
\tbl{Summary statistics of the sample stocks and their returns in Shanghai and Shenzhen A-Share markets. The information includes the number of stocks, number of records, the mean, standard deviation, maximum, minimum, skewness and kurtosis of the returns in both markets.}
{\resizebox{1.0\linewidth}{!}{
{\begin{tabular}{lccrrrrrr}
\hline\hline
  & \multirow{3}*[2mm]{No. of stocks}& \multirow{3}*[2mm]{No. of records} & \multicolumn{6}{c}{Return $r_i(t)$}\\
\cline{4-9}
     &  &  & Mean &	Std.dev  &	Max. &  Min.& Skewness & Kurtosis \\
\hline
 Shanghai A-Share market & 181 & 643404  &	0.0002 & 0.0284 & 0.1041 & -0.1164 &  -0.0441  &	5.3819 \\%
 Shenzhen A-Share market & 181 & 639607  &	0.0002 & 0.0286 & 0.1042 &	-0.1163 & -0.0792  &	5.3434 \\%
\hline\hline
  \end{tabular}}}}
 \label{Descriptive statistics}
\end{minipage}
\end{center}
\end{table}

The return series of a certain stock $i$ is computed as $r_i(t) = \ln P_i(t) - \ln P_i(t-1)$, where $P_i(t)$ is the closing price for stock $i$ on the $t$-th day. The price returns for $181\times 2$ stocks are calculated, and the effects of corporate actions are eliminated, for instance the cash dividend, the bonus share, and the rights issue. The summary statistics of the sample stocks and their price returns are listed in Table~\ref{Descriptive statistics}, including the number of stocks, the number of records, the mean, standard deviation, maximum, minimum, skewness and kurtosis of the returns in the two markets. The number of daily records of sample stocks for Shanghai and Shenzhen A-Share markets are 643,404 and 639,607 respectively. The values of the mean, standard deviation, maximum and minimum of the daily returns are very similar for the two markets. Returns in both markets show negative skewness and leptokurtosis, which have been widely observed in stock markets \citep{Mandelbrot-1963-JB,Cont-2001-QF}.

\section{Methods}

For a certain daily point $t$, a correlation matrix is calculated by the Pearson correlation coefficient estimator using the returns series in the window $\{t-\delta t+1,\ldots,t\}$, and based on which a stock network is constructed by the MST method, see details in network construction in Methods. Ten categories of portfolios are selected respectively from a set of 10\% most peripheral and central stocks in the MST graph, in which the centrality/peripherality is measured by five parameters capturing network topology: degree, betweenness centrality, distance on degree criterion, distance on correlation criterion and distance on distance criterion. For more details, see portfolio selection in Methods. The selected portfolios are used for investment in the following horizon $\{t+1,\ldots,t+\Delta t\}$, with a equal weight for each selected stock, following \citep{DeMiguel-Garlappi-Uppal-2009-RFS} in which 1/N portfolio strategy is proven to be more efficient than the mean-variance model. The investment returns of chosen portfolios are calculated under nine combinations of market conditions in the selection and investment horizons. The market condition includes drawup (U), drawdown (D) and stable (S) status identified by trading day criterion, amplitude criterion, "OR" criterion, and "AND" criterion. Please refer to the descriptions about identification of market conditions in Methods. Since the results obtained from three criterions are quantitatively similar, we mainly introduce our research results based on the trading day criterion. The data point then skips to $t+\varphi$, and the same portfolio strategy is adopted by selecting the portfolio in the window (horizon) $\{t+\varphi-\delta t+1,\ldots,t+\varphi \}$, and using the selected portfolio for investment in the horizon $\{t+\varphi+1,\ldots,t+\varphi+\Delta t\}$. The investment returns of 10 categories of selected portfolios are calculated under nine combinations of market conditions in the two horizons, and the optimal portfolio is found out evaluated by their average performances over different moving windows.

A suitable choice of $\delta t$ and $\varphi$ will indeed make the network captures the information of original data as much as possible. The larger $\delta t$ is and the smaller $\varphi$ is, the more stable the network structure is, and the more the market information is filtered out. On the contrary, the network structure is more volatile and unauthentic \citep{Strogatz-2001-Nature,Krings-Karsai-Bernhardsson-Blondel-Saramaki-2012-EPJDS}, though the temporary fluctuations can be easily noticed. Many studies have revealed that in order to ensure stocks have enough number of trading days to be statistically significant, $\delta t$ should be larger than the number of sample stocks $N=181$ \citep{Ledoit-Wolf-2004-JMA,Mardia-Kent-Bibby-1979}. By careful observation and precise calculation, we choose $\delta t=10$ months ($\approx 200$ days) and $\varphi=1$ month($\approx 20$ days), thus we have 161 daily points used for portfolio investments in total. The determination of the optimal values of the parameter $\varphi$ and the size of investment horizon $\Delta t$ will be described in detail in network construction and determination of investment horizon in Methods respectively.

\subsection{Network construction based on MST method}

Denote $r_i(t)$ and $r_j(t)$ as the logarithmic returns of stock $i$ and $j$, the Pearson correlation coefficient between their return series is given by
\begin{equation}
\rho(i,j)=\frac{E[r_i r_j]-E[r_i] E[r_j]}{\sqrt{(E[r_i^2]- E[r_i]^2)(E[r_j^2]- E[r_j]^2)}},\label{equ_correlation_coefficient}
\end{equation}
where $E[\cdot]$ represents the mathematical expectation of the sequence over time $t$.
Before the construction of MST graph, the correlation coefficient is converted into the distance between stock $i$ and $j$ by the following equation
\begin{equation}
d(i,j)=\sqrt{2(1-\rho(i,j))},\label{equ_distance}
\end{equation}
The distance $d(i,j)$ ranges from 0 to 2, and a small distance corresponds to a large correlation coefficient. For the $181$ sample stocks respectively in the Shanghai and Shenzhen A-Share markets, a distance matrix with $181\times 181$ elements is obtained for each market. The estimation of the correlation matrix has unavoidably associated with a statistical uncertainty, which is due to the finite length of the return series as well as noise.

We choose the MST method to filter out the network graph in each window so as to eliminate the redundancies and noise, and meanwhile, maintain significant links. In constructing the minimum spanning tree, we effectively reduce the information space from $n(n-1)/2$ correlation coefficients to $n-1$ tree edges, in other words, compress the amount of information dramatically. The procedure to build the MST network can be carried out as follows: First, arrange the distances between all pairs of stocks in an ascending order, then start by matching the nearest nodes, and continue matching following the ordered list if and only if the graph obtained after the matching is still a tree. Edges maximizing the sum of the correlations over the connections in the tree are more likely to stay in this method. Many researches like references \citep{Mantegna-1999-EPJB,Kim-Lee-Kahng-2002-JPSJ,Onnela-Kaski-Kertesz-2004-EPJB,Lee-Lee-Hong-2007-CPC,Micciche-Bonanno-Lillo-Mantegna-2003-PA} have used the MST model to filter networks. Given the data used in our study, we choose Prim algorithm to build our network.

There exists a close relationship between correlation coefficient matrices and MST distance matrices. To investigate their relationship, we calculate the Pearson linear correlation coefficients between the value of mean, variance, skewness and kurtosis of the elements in both matrices. Generally speaking, mean value of the elements in the two matrices are anti-correlated, and there exist similar feature for skewness. Variance, together with kurtosis of the elements in two matrices are positively correlated. These characteristics can be expected in view of how distances are constructed from the correlation coefficients. To confirm that, we provide the Pearson linear correlation coefficients of these variables in table~\ref{MST;original;compare}. We find that all variables of the elements in the two matrices are strongly correlated except kurtosis, showing that most of market information has been extracted by our MST network. In general, with a suitable choice of $\varphi$, the relationship between correlation coefficient matrices and MST distance matrices will be enhanced. With a series of tests, we find that $\varphi= 1$ month turns out to be the optimal choice.

\begin{table}
\begin{center}
\begin{minipage}[c]{1\linewidth}
\tbl{Pearson linear correlation coefficients between four pairs of variables. The four variables include the value of mean, variance, skewness and kurtosis of elements in correlation coefficient matrices and distance matrices for the two markets. }
{\begin{tabular}{lcrrrcrrr}
  \hline\hline
 & mean & variance &	skewness &	kurtosis\\
 \hline
 Shanghai A-Share market & -0.9849 & 0.7971 & -0.7302 &	0.2813\\%
 Shenzhen A-Share market & -0.9847 & 0.8011 & -0.8624 &	0.2584\\%
  \hline\hline
  \end{tabular}}
 \label{MST;original;compare}
\end{minipage}
\end{center}
\end{table}

\subsection{Portfolio selection based on topological parameters}

Five parameters are used to measure the centrality and peripherality of nodes in portfolio selection: I degree, II betweenness centrality, III distance on degree criterion, IV distance on correlation criterion and V distance on distance criterion. Here, we present a brief introduction of these parameters.

I. Degree $K$, the number of neighbor nodes connected to a node. The larger the \textit{K} is, the more the edges that are associated with this node.

II. Betweenness centrality $C$, reflecting the contribution of a node to the connectivity of the network. Denote $V$ as the set of nodes in the network, for nodes $i$ and $j$, $C$ of a node can be calculated as
\begin{equation}
C=\sum_{i,j\in V} \frac{\sigma_{ij}(V)}{\sigma_{ij}},
\end{equation}
where  $\sigma_{ij}$ is the number of shortest routes from node $i$ to node $j$, $\sigma_{ij}(V)$ is a subsector of $\sigma_{ij}$  whose routes pass through this node.

Distance refers to the smallest length from a node to the central node of the network. Here, three types of definitions of central node are introduced to reduce the error caused by a single method. Therefore three types of distances are described here.

III. Distance on degree criterion $D_{degree}$, central node is the one that has the largest degree.

IV. Distance on correlation criterion $D_{correlation}$, central node is the one with the highest value of the sum of correlation coefficients with its neighbors;

V. Distance on distance criterion $D_{distance}$, central node is the one that produces the lowest value for the mean distance.

We use the parameters defined above to select the portfolios. Nodes with the largest 10\% of degree or betweenness centrality are chosen to be in the central portfolio, and nodes whose degree equals to 1 or betweenness centrality equals to 0 are chosen to be in the peripheral portfolio. Similarly, we define the nodes ranking the top 10\% of distance as the stocks of the peripheral portfolios, and the bottom 10\%  as the stocks of the central portfolios. The difference in the definitions results from a simple reason: For an MST network, the number of peripheral nodes (i.e., leaf nodes of a network), whose degree equals to 1 and betweenness centrality equals to 0, is much larger than 10\%  of the total nodes. We need to mention that it makes no difference to our results if we select randomly from these peripheral nodes so as to equal to the number in each portfolio.

The central portfolios and peripheral portfolios represent two opposite side of correlation and agglomeration. Generally speaking, central stocks play a vital role in the market and impose strong influence on other stocks. While the correlations between peripheral stocks are weak and contain much more noise than central stocks. We have learned in our study that the two kinds of portfolios have their own features under different market conditions.

\subsection{Determination of investment horizon}

In this part we will discuss the optimal choice of the length of investment horizons $\Delta t$. In general, the length of investment horizon cannot be too long, otherwise the topological properties of the network will change and the selected central or peripheral portfolios will change accordingly. On the other hand, the length of investment horizons cannot be too short, or the returns will be greatly influenced by market noises or exogenous events. Here, we compare the profits gained in different time horizons, namely 1 month, 5 months, 10 months and 15 months.

Sharpe ratio, defined as the ratio of expected value of the excess returns to its standard deviation \citep{Sharpe-1994-JPM}, is widely used to evaluate the performance of portfolios practically \citep{Jagannathan-Ma-2003,Scholz-2007-JAM}. Table~\ref{timehorizon} shows Sharpe ratios of portfolios with different investment horizons, in which there is no specular length that could maximize Sharpe ratio in all circumstances. What's more, one-way ANOVA is used to test the equality of excess returns gained in different length of investment horizons. In table~\ref{timehorizon}, the $p$-values for different horizons are all insignificant, which indicates the choice of horizon's length doesn't affect the portfolio returns. In order to match the length of selection horizons and investment horizons and facilitate our identification of the market conditions, we choose $\Delta t =10$ months for portfolio investment.

\begin{table}
\begin{center}
\begin{minipage}[c]{\linewidth}
\tbl{Sharpe ratio of portfolios for different lengths of investment horizons and ANOVA test of portfolios' excess returns. Portfolios listed include central and peripheral portfolios selected with respect to $K$, $C$, $D_{degree}$, $D_{correlation}$ and $D_{distance}$ for the Shanghai and Shenzhen A-Share markets.}
{\resizebox{1.0\linewidth}{!}{
{\begin{tabular}{llrrrrrcrrrrr}
  \hline\hline
     && \multicolumn{5}{c}{Shanghai A-Share market} & & \multicolumn{5}{c}{Shenzhen A-Share market} \\
  \cline{3-7}  \cline{9-13}
    && 1 month & 5 month & 10 month & 15 month & $p$-value & &  1 month & 5 month & 10 month & 15 month & $p$-value \\
  \hline
  \multirow{2}{*}{$K$} & peripheral	 & -0.0113 	& 0.0024 	& -0.0005 	& -0.0023 	& 0.9968 	& & -0.0084 & 0.0025 	& -0.0007 	& -0.0028 	& 0.9964\\
  & central   & 0.0455 	& 0.0112 	& 0.0043 	& 0.0002 	& 0.9652 	& & 0.0444 	& 0.0120 	& 0.0061 	& 0.0030 	& 0.9892\\
  \multirow{2}{*}{$C$}& peripheral   & -0.0113 	& 0.0024 	& -0.0005 	& -0.0023 	& 0.9968 	& & -0.0084 & 0.0025 	& -0.0007 	& -0.0028 	& 0.9964\\
  & central   & 0.0387 	& 0.0058 	& 0.0016 	& -0.0015 	& 0.9635 	& & 0.0281 	& 0.0074 	& 0.0028 	& 0.0002 	& 0.9915\\
  \multirow{2}{*}{$D_{degree}$}& peripheral	 & -0.0379 	& -0.0488 	& -0.0724 	& -0.1228 	 & 0.9978 	& & -0.0281 & -0.0048 	& -0.0054 	& -0.0072 	& 0.9954\\
  & central   & 0.0349 	& 0.0386 	& 0.0446 	& 0.0367 	& 0.9941 	& & 0.0389 	& 0.0155 	& 0.0069 	& 0.0033 	& 0.9923\\
  \multirow{2}{*}{$D_{correlation}$}& peripheral	 & -0.0317 	& -0.0109 	& -0.0079 	& -0.0084 	& 0.9991 	& & -0.0385 & -0.0043 	& -0.0052 	& -0.0067 	& 0.9911\\
  & central   & 0.0291 	& 0.0061 	& 0.0041 	& 0.0016 	& 0.9958 	& & 0.0450 	& 0.0121 	& 0.0067 	& 0.0033 	& 0.9912\\
  \multirow{2}{*}{$D_{distance}$}& peripheral	 & -0.0351 	& -0.0091 	& -0.0067 	& -0.0079 	& 0.9979 	& & -0.0356 & -0.0024 	& -0.0048 	& -0.0068 	& 0.9876\\
  & centeral   & 0.0195 	& 0.0075 	& 0.0038 	& 0.0010 	& 0.9984 	& & 0.0268 	& 0.0107 	& 0.0052 	& 0.0020 	& 0.9966\\
  \hline\hline
  \end{tabular}}}}
\label{timehorizon}
\end{minipage}
\end{center}
\end{table}

\subsection{Identification of market conditions}

Market conditions, which describes the general trend of the market index over a specific horizon, are measured by using four criteria: I trading day criterion, II amplitude criterion, III "OR" criterion and IV "AND" criterion. The market indices corresponding to the Shanghai A-Share market and the Shenzhen A-Share market include the Shanghai A-Share Index and the Shenzhen A-Share Index.

I. Trading day criterion. The ratio  $r_d$ of the number of days with rising index to the total number of trading days in a specific time window is given by,
\begin{equation}
r_d=\frac{{N_i}^{+}}{N_i},
\end{equation}
where ${N_i}^{+}$ is the number of days in which the closing price is larger than that of the previous day and $N_i$ is the total number of trading days in the $i$-th time window. The ratio $r_d$ ranges from 0 to 1, and a large value of $r_d$ represents a drawup condition, while a small value of $r_d$ represents a drawndown condition. With the thresholds $\theta_{+}$ and $\theta_{-}$, we identify a drawup condition if $r_d>\theta_{+}$, a drawdown condition if $r_d<\theta_{-}$, and a stable condition if  $\theta_{-}\leq r_d \leq \theta_{+}$.

II. Amplitude criterion. The ratio $r_f$ of the sum of the amplitudes of the trading days with rising index to the sum of the amplitudes of the total trading days in a specific time window is given by,
\begin{equation}
r_f=\frac{\sum_{t\in {T_i}^{+}} |P(t)-P(t-1)|}{\sum_{t\in T_i} |P(t)-P(t-1)|},
\end{equation}
where ${T_i}^{+}$ is the set of days in which the closing price is larger than that of the previous day, $T_i$ is the set of all the trading days in the $i$-th time window, and $P(t)$ is the closing price on the $t$-th day. Similarly, with the thresholds $\theta_{+}$ and $\theta_{-}$, we identify a drawup condition if $r_f>\theta_{+}$, a drawdown condition if $r_f<\theta_{-}$, and a stable condition if $\theta_{-}\leq r_f \leq \theta_{+}$.

III. "OR" criterion, we identify a drawup condition if $r_d>\theta_{+}$ or $r_f>\theta_{+}$, and a drawdown condition if $r_d<\theta_{-}$ or $r_f<\theta_{-}$. A stable condition is identified if $\theta_{-} \leq r_d \leq \theta_{+}$ and $\theta_{-} \leq r_f \leq \theta_{+}$. Situations like $r_d >\theta_{+}$ and $r_f <\theta_{-}$, or $r_f >\theta_{+}$ and $r_d <\theta_{-}$ do not exist under the thresholds we choose.

IV. "AND" criterion, we identify a drawup condition if $r_d>\theta_{+}$  and $r_f>\theta_{+}$, and a drawdown condition if $r_d<\theta_{-}$  and $r_f<\theta_{-}$. A stable condition is identified if $\theta_{-} \leq r_d \leq \theta_{+}$ or $\theta_{-} \leq r_f \leq \theta_{+}$. Situations like $r_d >\theta_{+}$ and $r_f <\theta_{-}$, or $r_f >\theta_{+}$ and $r_d <\theta_{-}$ do not exist under the thresholds we choose.

\begin{figure}
\begin{center}
\begin{minipage}{11cm}
\resizebox*{11cm}{!}{\includegraphics{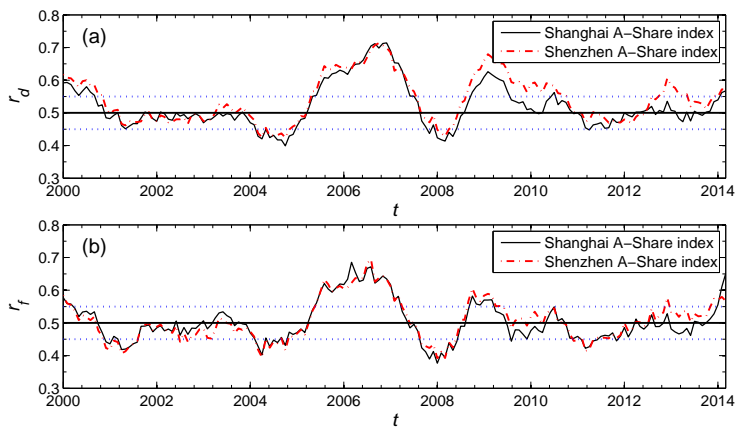}}
\end{minipage}
\end{center}
\caption{\label{rdrf} Ratios $\textit{r}_{\textit{d}}$ and $\textit{r}_{\textit{f}}$ as a function of time $t$, together with the thresholds $\theta_{+}=0.55$ and $\theta_{-}=0.45$ depicted by blue dotted lines.}
\end{figure}

The ratios $r_d$  and  $r_f$  over time are shown in figure~\ref{rdrf}, each point on the curve is measured by using the market indices in a 10-month horizon following this point. In the figure, the patterns of $r_d$ and $r_f$ show some differences, which cause slight distinctions in the identification of market conditions. In our study, we choose $\theta_{+}=0.55$, $\theta_{-}=0.45$ as the thresholds. Theoretically, other choices of thresholds will work in our study. For a larger value of $\theta_{+}$ and a smaller value of $\theta_{-}$, the number of samples of drawup or drawdown conditions is not statistically sufficient for the lack of data. For a smaller value of $\theta_{+}$ and a larger value of $\theta_{-}$, the drawup and drawdown conditions cannot be distinctly identified. Other suitable choices of thresholds around $\theta_{+}=0.55$, $\theta_{-}=0.45$ have also been studied and the results do not change significantly.

\begin{figure}
\begin{center}
\begin{minipage}{11cm}
\resizebox*{11cm}{!}{\includegraphics{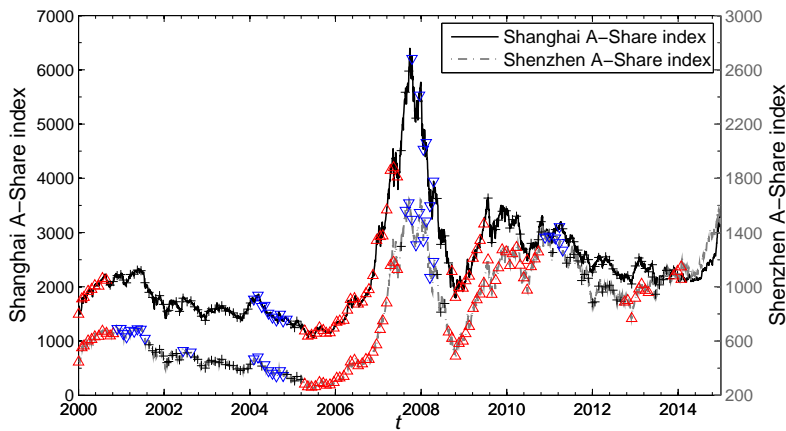}}
\end{minipage}
\end{center}
\caption{\label{trend} Market conditions identified based on trading day criterion for the Shanghai and Shenzhen A-Share markets, including drawup (upper triangle), drawdown (lower triangle) and stable (cross) conditions.}
\end{figure}

Market conditions identified based on the trading day criterion for the Shanghai and Shenzhen A-Share markets are shown in figure~\ref{trend}. The upper triangle indicates a drawup horizon from the current time to 10 months later, and similarly the lower triangle indicates a drawdown horizon and the cross symbol indicates a stable horizon. It seems that most of the drawup and drawdown conditions identified in our study are proper and sustainable in both markets.

For each time window we get three possible market conditions, we thus get nine combinations of market conditions in selection and the following investment horizons: Drawup in both selection horizon and investment horizon (UU), drawup in the selection horizon and stable in the investment horizon (US), drawup in the selection horizon and drawdown in the investment horizon (UD), stable in the selection horizon and drawup in the investment horizon (SU), stable in both selection and investment horizons (SS), stable in the selection horizon and drawdown in the investment horizon (SD), drawdown in the selection horizon and drawup in the investment horizon (DU), drawdown in the selection horizon and stable trend in the investment horizon (DS), drawdown in both selection and investment horizons (DD).

\section{Results}

\subsection{Evolution of network structures}

\begin{figure}
\begin{center}
\begin{minipage}{11cm}
\resizebox*{11cm}{!}{\includegraphics{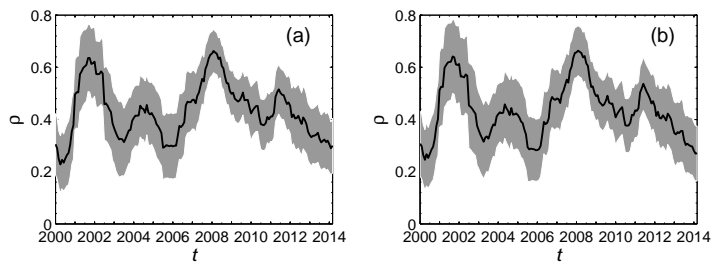}}
\end{minipage}
\end{center}
\caption{\label{Correlation Coefficient} Evolution of average correlation coefficients in Shanghai A-Share market (a) and Shenzhen A-Share market (b). The average correlation coefficients are shown by the black solid lines in the center, and correlation coefficients ranging within a standard deviation are shown in the grey area.}
\end{figure}

Practically, the network structure is evolving over time and changes remarkably during crises. Some evolutionary characteristics of the market can be found from the topological parameters of the networks. First, the average correlation coefficients among all stocks reflect the overall connections of the spanning tree, which are shown in figure~\ref{Correlation Coefficient}. To learn more details about the evolution of correlation coefficients in each window, the correlation coefficients ranging within a standard deviation are shown. The average correlation coefficients rise sharply in the periods of market crashes in 2001 and 2008. As the market recovers, the average correlation coefficients decline correspondingly. This finding is consistent with the conclusions illustrated by previous researches, indicating that the connection between stocks will be enhanced during the crisis \citep{Onnela-Chakraborti-Kaski-Kertesz-Kanto-2003-PRE,Drozdz-Grummer-Gorski-Ruf-Speth-2000-pa}.

\begin{figure}
\begin{center}
\begin{minipage}{8cm}
\resizebox*{8cm}{!}{\includegraphics{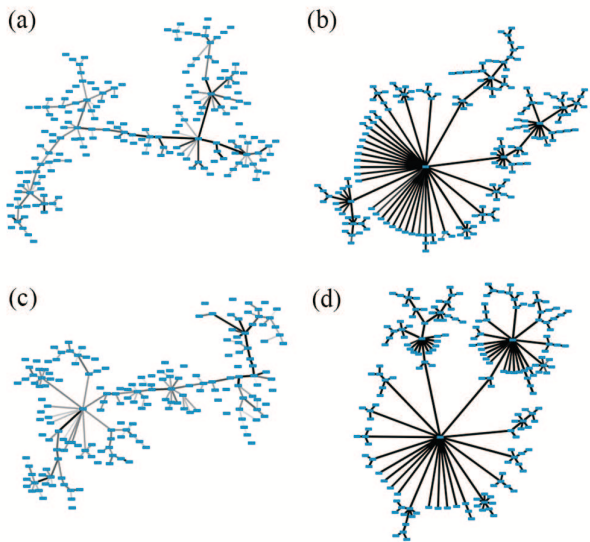}}
\end{minipage}
\end{center}
\caption{\label{MST tree;sh;sz}(a) MST network for the Shanghai A-Share market in the period from January 1, 2000 to October 31, 2000; (b) MST network for the Shanghai A-Share market in the period from January 1, 2008 to October 31, 2008; (c) MST network for the Shenzhen A-Share market in the period from January 1, 2000 to October 31, 2000; (d) MST network for the Shenzhen A-Share market in the period from January 1, 2008 to October 31, 2008. Distances between stocks are indicated by line width: A thicker line represents a shorter distance while a thinner line represents a longer distance.}
\end{figure}

To better understand the variance of network structure over time $t$, we choose two typical examples of MST networks begin at January 1, 2000 and January 1, 2008, when stock prices are stable in the former window while volatile in the latter. Their network structures are shown in figure~\ref{MST tree;sh;sz}. We can see that the distances between stocks are much smaller, and the corresponding network shrinks to a large extent during the 2008 market crisis.

More specific than correlation coefficients, other parameters namely degree, betweenness centrality, distance on degree criterion, distance on correlation criterion and distance on distance criterion, can tell us more information about the stocks and networks. These parameters are used for selecting portfolios in our later study.

\begin{figure}
\begin{center}
\begin{minipage}{11cm}
\resizebox*{11cm}{!}{\includegraphics{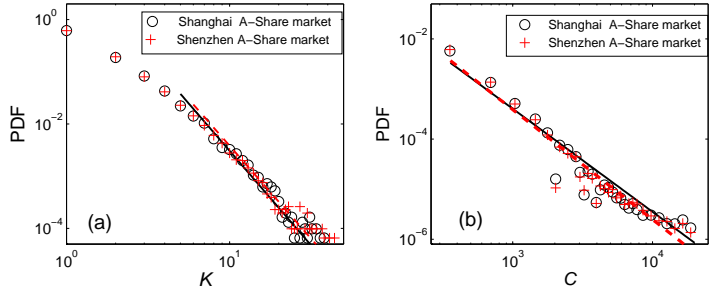}}
\end{minipage}
\end{center}
\caption{\label{degree;distribution} PDFs of degree $K$ and betweenness centrality $C$ in the Shanghai A-Share market (a) and the Shenzhen A-Share market (b). The fitted lines with exponents estimated by a maximum likelihood estimation method proposed by Newman are for the Shanghai A-Share (black solid line) and Shenzhen A-Share markets (red dashed line).}
\end{figure}

\begin{figure}
\begin{center}
\begin{minipage}{18cm}
\resizebox*{18cm}{!}{\includegraphics{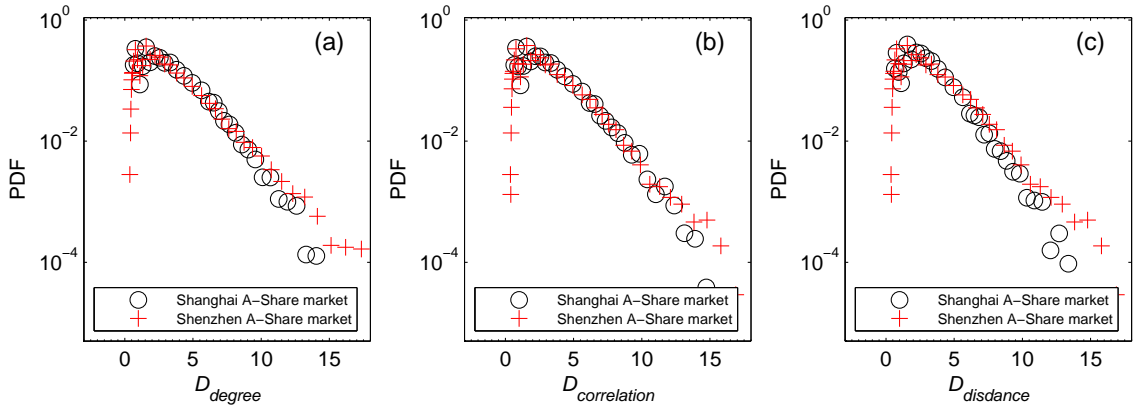}}
\end{minipage}
\end{center}
\caption{\label{distance;distribution} (a) PDF of distance on degree criterion $D_{degree}$, (b) PDF of distance on correlation criterion $D_{correlation}$ and (c) PDF of distance on distance criterion $D_{distance}$ in the Shanghai A-Share and Shenzhen A-Share markets.}
\end{figure}

The parameter degree $K$ describes the relationship between a stock and its neighbors. Many studies have found that degree in empirical networks follows a power-law distribution \citep{Newman-2001a-PRE,Strogatz-2001-Nature,Lee-Lee-Hong-2007-CPC}, where the distribution of $K$ has a formula $P(K)\sim K^{-\alpha}$. Here, we observe a power-law behavior for the probability density function (PDF) of $K$ in figure~\ref{degree;distribution}. A maximum likelihood estimation method proposed by Newman is used to fit the distributions \citep{Clauset-Shalizi-Newman-2009-SIAMR}. The exponents $\alpha=3.6202$, $\alpha=3.5597$ are estimated for $P(K)$ in Shanghai and Shenzhen A-Share markets respectively. Another parameter betweenness centrality $C$ reflects the contribution of a stock to the connectivity of the whole network, resembling degree to some extent. PDF of $C$, which is also shown in figure~\ref{degree;distribution}, also shows a power-law behavior. The exponents $\alpha=2.0562$, $\alpha=2.1927$ are estimated for $P(C)$ in Shanghai and Shenzhen A-Share markets. Results in the figure show the scale free property of MST networks in the two markets, and also the intense connection between stocks existing in a small number of central stocks, whose volatility might have a great impact on its neighbors.

Distance, known as the total length from a node to the central node of the network, has three categories according to the choice of central node in our study, namely, distance on degree criterion $D_{degree}$, distance on correlation criterion $D_{correlation}$ and distance on distance criterion $D_{distance}$. Their PDFs are shown in figure~\ref{distance;distribution}. Although the three categories of distances vary from each other, their distributions as well as the stocks selected by them share similar behavior. As can be seen from figure~\ref{distance;distribution}, few stocks are very distant from the central node, while most stocks have a relatively short distance with the central node. In addition, note that the MST networks change over time, therefore its central node also changes.

\subsection{Comparison between portfolio strategies under different market conditions}

\begin{table}
\begin{center}
\begin{minipage}[c]{1\linewidth}
\tbl{ Excess returns of central and peripheral portfolios in the Shanghai A-Share market are compared based on one-way ANOVA. Central and peripheral portfolios are selected based on five parameters, i.e., $K$, $C$, $D_{degree}$, $D_{correlation}$ and $D_{correlation}$, and their excess returns are calculated under different combinations of market conditions based on trading day criterion. The listed variables include the number of samples (Num), $f$-value and $p$-value of one-way ANOVA, excess returns of central and peripheral portfolios under each combination of market conditions. Results which are not significant or calculated with less than 11 samples are not shown. ( *indicates significance at 10\% level, **indicates significance at 5\% level ).}
{\begin{tabular}{lccllccc}
  \hline\hline
  \multirow{3}*[2mm]{Parameter}& \multirow{3}*[2mm]{Market condition} & \multirow{3}*[2mm]{Num} & \multirow{3}*[2mm]{$f$-value} & \multirow{3}*[2mm]{$p$-value} & & \multicolumn{2}{c}{excess returns}\\
  \cline{7-8}
     &  &  &  &  & & central & peripheral \\
  \hline
   \multirow{4}{*}{$K$} & US   & 24  & 15.17  & 0.00** & & \textbf{0.04}  & 0.00 \\
   & SS                      & 74  & 6.75   & 0.01** & & \textbf{0.02}  & 0.00 \\
   & SD                      & 11  & 5.79   & 0.03** & & \textbf{0.02}  & 0.00 \\
   & DU                      & 11  & 3.87   & 0.06*  & & \textbf{0.03}  & 0.00 \\
   \cline{1-8}
   \multirow{5}{*}{$D_{degree}$} & US  & 24  & 5.69   & 0.02** & & \textbf{0.02}  & -0.03 \\
   & SU                      & 12  & 4.27   & 0.05*  & & \textbf{0.02}  & -0.04 \\
   & SS                      & 74  & 21.39  & 0.00** & & \textbf{0.02}  & -0.03 \\
   & SD                      & 11  & 3.57   & 0.07*  & & -0.02  & \textbf{0.04} \\
   & DU                      & 11  & 53.08  & 0.00** & & \textbf{0.05}  & -0.08 \\
   \cline{1-8}
   \multirow{4}{*}{$D_{correlation}$} & US & 24  & 5.33   & 0.03** & & \textbf{0.02}  & -0.03 \\
   & SU                      & 12  & 4.27   & 0.05*  & & \textbf{0.02}  & -0.04 \\
   & SS                      & 74  & 20.82  & 0.00** & & \textbf{0.02}  & -0.03 \\
   & DU                      & 11  & 52.77  & 0.00** & & \textbf{0.05}  & -0.08 \\
   \cline{1-8}
   \multirow{5}{*}{$D_{distance}$} & US  & 24  & 3.98   & 0.05*  & & \textbf{0.01}  & -0.03 \\
   & SU                      & 12  & 4.18   & 0.05*  & & \textbf{0.01}  & -0.03 \\
   & SS                      & 74  & 19.81  & 0.00** & & \textbf{0.03}  & -0.03 \\
   & SD                      & 11  & 4.36   & 0.05*  & & -0.02  & \textbf{0.04} \\
   & DU                      & 11  & 56.84  & 0.00** & & \textbf{0.05}  & -0.08 \\
  \hline\hline
  \end{tabular}}
\label{returnComparison;sh}
\end{minipage}
\end{center}
\end{table}

In this part, we compare the returns of central and peripheral portfolios and find the optimal portfolio among them. The portfolios are selected by using five parameters, i.e., $K$, $C$, $D_{degree}$, $D_{correlation}$ and $D_{correlation}$, in the selection horizon, and the returns of the selected portfolios are calculated in the following investment horizon. The length of investment horizon is  set to be 10 months as we discuss in Determination of investment horizon in Methods. In this paper, the selection horizon lasts from January 1, 2000 to February 31, 2014, and the investment horizon lasts from November 1, 2000 to December 31, 2014,  thus we have 161 daily points used for portfolio investments in total.

We calculate the returns of central portfolios and those of peripheral portfolios, and we use the returns of random portfolios as a benchmark. Random portfolio is defined as a random selected portfolio containing $10\%$ of the total stocks. We first classify the samples of returns of selected portfolios and random portfolios into groups according to nine combinations of market conditions identified using thresholds $\theta_{+}=0.55$, $\theta_{-}=0.45$ based on trading day criterion. For each combination of market conditions, we calculate the average return of each individual stock in the group of selected portfolios and the average return of each individual stock in the group of random portfolios. The difference between the average returns of selected portfolios and random portfolios is defined as excess return. Furthermore, one-way ANOVA is used to test the equality of the excess returns between central portfolios and peripheral portfolios under the same market condition. Null hypothesis of one-way ANOVA, that the excess returns of central portfolios and peripheral portfolios are equal, are tested under a specific significance level. If the null hypothesis is rejected, the excess returns of the two portfolios are significantly different. If the null hypothesis cannot be rejected, there is no significant difference between excess returns of central portfolios and those of peripheral portfolios.

\begin{table}
\begin{center}
\begin{minipage}[c]{1\linewidth}
\tbl{ Excess returns of central and peripheral portfolios in the Shenzhen A-Share market are compared based on one-way ANOVA. Central and peripheral portfolios are selected based on five parameters, i.e., $K$, $C$, $D_{degree}$, $D_{correlation}$ and $D_{correlation}$, and their excess returns are calculated under different combinations of market conditions based on trading day criterion. The listed variables include the number of samples (Num), $f$-value and $p$-value of one-way ANOVA, excess returns of central and peripheral portfolios under each combination of market conditions. Results which are not significant or calculated with less than 11 samples are not shown.}
{\begin{tabular}{lccllccc}
  \hline\hline
 \multirow{3}*[2mm]{Parameter}& \multirow{3}*[2mm]{Market condition} & \multirow{3}*[2mm]{Num} & \multirow{3}*[2mm]{$f$-value} & \multirow{3}*[2mm]{$p$-value} & & \multicolumn{2}{c}{excess returns}\\
  \cline{7-8}
     &  &  &  &  & & central & peripheral \\
  \hline
  \multirow{4}{*}{$K$} & UU & 37  & 2.87  & 0.09*  & & \textbf{0.02}  & 0.00 \\
  & US                    & 32  & 9.75  & 0.00** & & \textbf{0.03}  & -0.01 \\
  & SU                    & 20  & 3.93  & 0.05*  & & \textbf{0.03}  & 0.00 \\
  & SS                    & 50  & 19.83 & 0.00** & & \textbf{0.02}  & -0.01 \\
  \cline{1-8}
  \multirow{1}{*}{$C$} & SS & 50  & 8.00  & 0.01**  & & \textbf{0.01}  & -0.01 \\
  \cline{1-8}
  \multirow{3}{*}{$D_{degree}$} & UU & 37  & 10.35 & 0.00**  & & \textbf{0.02}  & -0.04 \\
  & SU                    & 20  & 20.66 & 0.00**  & & \textbf{0.05}  & -0.05 \\
  & SS                    & 50  & 8.50  & 0.00**  & & \textbf{0.02}  & -0.01 \\
  \cline{1-8}
  \multirow{4}{*}{$D_{correlation}$} & UU & 37  & 3.06  & 0.08*   & & \textbf{0.00}  & -0.03 \\
  & US	                  & 32	& 2.82  & 0.10*   & & \textbf{0.03}  & -0.01 \\
  & SU                    & 20  & 20.66 & 0.00**  & & \textbf{0.05}  & -0.05 \\
  & SS                    & 50  & 11.91 & 0.00**  & & \textbf{0.03}  & -0.02 \\
  \cline{1-8}
  \multirow{3}{*}{$D_{distance}$} & UU & 37	& 6.62  & 0.01**  & & \textbf{0.01}  & -0.03 \\
  & SU                    & 20  & 4.76  & 0.04**  & & \textbf{0.02}  & -0.04 \\
  & SS                    & 50  & 10.63 & 0.00**  & & \textbf{0.02}  & -0.02 \\
   \hline\hline
  \end{tabular}}
\label{returnComparison;sz}
\end{minipage}
\end{center}
\end{table}

The results of one-way ANOVA test and the excess returns of central and peripheral portfolios for Shanghai and Shenzhen A-Share markets are listed in Tables \ref{returnComparison;sh} and \ref{returnComparison;sz} respectively. Note that if the sample number is less than 11 under a specific combination of market conditions, we do not show the result of this combination for the lack of data. In addition, if there is no significant difference between excess returns of central portfolios and peripheral portfolios, the results are also not shown.

In table~\ref{returnComparison;sh}, we can see from the $p$-value that the excess returns between central and peripheral portfolios of the listed groups are all significantly different at the 10\% level, and most of them are significantly different at the 5\% level. By comparing the excess returns of central and peripheral portfolios under all the listed combinations of market conditions, we find that central portfolios are more profitable except for two cases. More specifically, when the market is stable in the investment horizon, whether the market is stable or drawup in the selection horizon, the excess returns of central portfolios are significantly larger than those of peripheral portfolios. When the market is drawdown in the selection horizon and drawup in the investment horizon, or when the market is stable in the selection horizon and drawup in the investment horizon, the central portfolios outperform peripheral portfolios using $K$, $D_{degree}$, $D_{correlation}$ and $D_{distance}$ as parameter. Besides, when the market is stable in the selection horizon and drawdown in the investment horizon, peripheral portfolios gain more than central portfolios using $D_{degree}$ and $D_{distance}$ as parameter. However, under the same market condition while using $K$ as parameter, the optimal portfolio is the central one with a relatively small gap between the excess returns of the two portfolios.

In table~\ref{returnComparison;sz} for the Shenzhen A-Share market, we find the results are generally coincide with those for Shanghai A-Share market. Central portfolios outperform peripheral portfolios under every combination of market conditions when there exist significant difference between their returns. What's more, central portfolios selected by $K$,  $D_{degree}$, $D_{correlation}$ and $D_{distance}$ outperform peripheral portfolios when the market goes through a rising trend in both selection and investment horizons.

\begin{figure}
\begin{center}
\begin{minipage}{13cm}
\resizebox*{13cm}{!}{\includegraphics{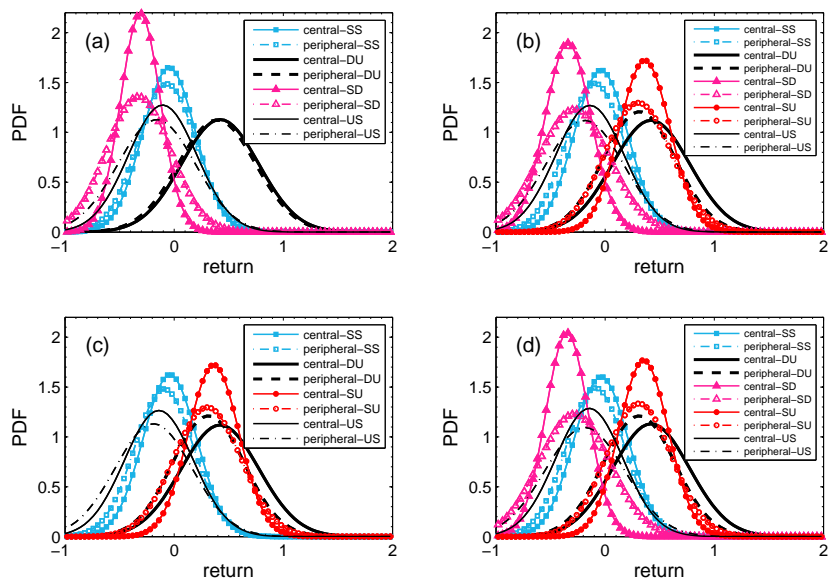}}
\end{minipage}
\end{center}
\caption{\label{frequency;return;sh} PDF of individual stock returns under each combination of market conditions based on trading day criterion in the Shanghai A-Share Market are plotted. Stock portfolios are selected with respect to degree $K$ (a), distance on degree criterion $D_{degree}$ (b), distance on correlation criterion $D_{correlation}$ (c), and distance on distance criterion $D_{distance}$ (d). }
\end{figure}

\begin{figure}
\begin{center}
\begin{minipage}{13cm}
\resizebox*{13cm}{!}{\includegraphics{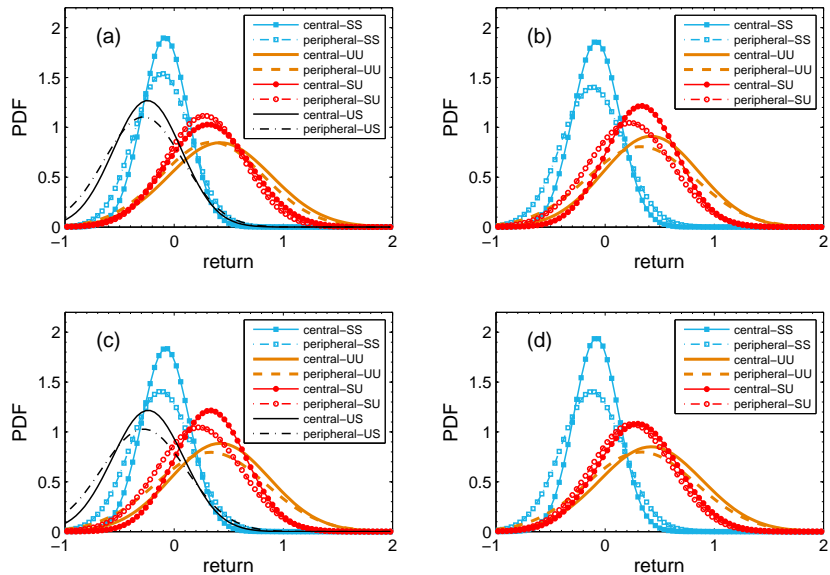}}
\end{minipage}
\end{center}
\caption{\label{frequency;return;sz} PDF of individual stock returns under each combination of market conditions based on trading day criterion in Shenzhen A-Share Market are plotted. Stock portfolios are selected with respect to degree $K$ (a), distance on degree criterion $D_{degree}$ (b), distance on correlation criterion $D_{correlation}$ (c), and distance on distance criterion $D_{distance}$ (d).}
\end{figure}

Here, we provide the PDF of the returns of individual stocks in central and peripheral portfolios for the Shanghai and Shenzhen A-Share markets in figures \ref{frequency;return;sh} and \ref{frequency;return;sz} respectively. In figure~\ref{frequency;return;sh}, under most market conditions, the center of distribution for central portfolios is on the right side of that for peripheral portfolios, indicating that the returns of central portfolios are on average larger than those of peripheral portfolios. However, when portfolios are selected with respect to $D_{degree}$ or $D_{distance}$, the center of distribution for peripheral portfolios is on the right side of that for central portfolios under market conditions of SD. These results are consistent with that from table~\ref{returnComparison;sh}. We can also learn from figure~ \ref{frequency;return;sh} that returns of central portfolios distribute at a relatively narrow range compared with returns of peripheral portfolios under most market conditions, showing the close connection among stocks in central portfolios.
Figure~\ref{frequency;return;sz} plots the PDFs of the returns of individual stocks for Shenzhen A-Share market. The center of distribution for central portfolios is on the right side of that for peripheral portfolios under all market conditions, indicating that the returns of central portfolios are on average larger than those of peripheral portfolios. Meanwhile, all the returns of central portfolios have narrow distributions compared with returns of peripheral portfolios, showing the diversification of returns of stocks in peripheral portfolios.

Similar test and comparison are carried out using amplitude criterion, "AND" criterion and "OR" criterion in identification of market conditions for both markets. Using amplitude criterion and "AND" criterion, the results are very similar to those using trading day criterion. For "OR" criterion, we find that if the market goes through a turnaround, falling in the selection horizon and rallying in the investment horizon, the central portfolios are proven to be more profitable. More comprehensively, extra efforts have been made for other values of thresholds in the identification of market conditions. Despite some statistical distinction, the conclusions we get show great similarities with the above, which in turn confirm the reliability of our conclusion.

We summarize and interpret our results in three aspects.
First, if the market rises in the investment horizon, the central portfolios should be the best choice. Specifically, if the market has a drawup trend in both selection and investment horizons, stocks in central portfolios will be more likely to rise due to their collective movements in rising, while stocks in peripheral portfolios may be too diversified to make profits. If the market rises in the investment horizon after declining, stocks in central portfolios are more likely to suffer losses in the selection horizon, and are more likely to rise in the investment horizon after hitting rock bottom.
Second, if the market is stable in the investment horizon and has just gone through a drawup or stable trend in the selection horizon, central portfolios are preferred. Since stocks in central portfolios are closely related, their prices move in a similar way. After the drawup or stable trend in the selection horizon, stocks in central portfolios tend to maintain the drawup or stable trend in the investment horizon. Returns of stocks in peripheral portfolios are too diversified in the stable investment horizon, some negative returns of individual stocks will be more likely to be in the portfolio.
Last, if the market falls after a period of stable fluctuation, peripheral portfolios are preferred so as to avoid risks. The diverse characteristic of peripheral portfolios is a good way to reduce risk and secure capital.

\subsection{Empirical test of optimal portfolio strategy}

We have compared the performances of central and peripheral portfolios under different combinations of market conditions. We now attempt to pick out the optimal portfolio strategy and apply it to make real investment based on an empirical test. We use the data from 2000 to 2010 to select the optimal portfolio under each specific combination of market conditions through a training process, and use the selected optimal strategy according to the current market condition to make investment based on the data from 2010 to 2014. The test is performed specifically as follows.

I.	Training to find the optimal portfolio strategy. The optimal portfolio strategy is picked out using the methods mentioned in the last section. We first select central and peripheral portfolios based on five parameters of the network in the selection horizon, and then invest these portfolios in the investment horizon. Excess returns of central and peripheral portfolios under each market condition, identified by one of the four criteria, are calculated and tested using one-way ANOVA. If the excess returns of central and peripheral portfolios are significantly different, portfolio with the higher excess returns is chosen as the optimal portfolio under the specific market condition. An optimal portfolio strategy comprises all the optimal portfolios under different combinations of market conditions.

II. Applying optimal strategy to investment. Before investment, market conditions in the investment horizon need to be predicted. Since the policy and economic environment have a great impact on Chinese stock markets, it's possible to make a general assessment of the prospect for the stock markets in the investment horizon based on the current market information. To simplify our study, we identify the market conditions in the investment horizon using empirical data, and in other words, our strategy performs well when the market condition in the investment horizon has a clear trend. Based on the identified combination of market conditions, the optimal portfolio of the optimal strategy picked out in step I is then selected and used for further investment. If the combination of market conditions do not appear, investment will not be made. The length of selection and investment horizons are 10 months, the same as in the previous section.

\begin{table}
\begin{center}
\begin{minipage}[c]{1\linewidth}
\tbl{Excess returns listed are gained by the optimal portfolio strategy for the Shanghai and Shenzhen A-Share markets. The optimal strategy is picked out from all the possible categories of portfolios selected with respect to five parameters, i.e., $K$, $C$, $D_{degree}$, $D_{correlation}$ and $D_{distance}$, based on four criteria, i.e., trading day criterion,  amplitude criterion, "OR" criterion and "AND" criterion. Excess returns in boldface are positive which have returns larger than the random strategy. }
{\begin{tabular}{lrrrrcrrrr}
  \hline\hline
  & \multicolumn{4}{c}{Shanghai A-Share market} & & \multicolumn{4}{c}{Shenzhen A-Share market}\\
  \cline{2-5} \cline{7-10}
    & trading day & amplitude & "OR"  & "AND"  & & trading day  & amplitude &  "OR"  & "AND" \\
     &  criterion & criterion	&  criterion & criterion & &   criterion	& criterion &  criterion & criterion\\
 \hline
   $K$   & \textbf{0.0169}  & \textbf{0.0065} 	& \textbf{0.0072} 	& \textbf{0.0156} & &  \textbf{0.0131} 	& \textbf{0.0113} & \textbf{0.0190} 	& \textbf{0.0238} \\
   $C$  & -0.0016 	& -0.0003           & \textbf{0.0129}           & -0.0033         & & -0.0030       	& -0.0135   & -0.0118            & -0.0049 \\
   $D_{degree}$ & \textbf{0.0345} 	& \textbf{0.0025} 	& \textbf{0.0072} 	& -0.1290  & &  \textbf{0.0303} 	& \textbf{0.0381} & -0.0093 	& \textbf{0.0529}\\
   $D_{correlation}$  & \textbf{0.0385} & \textbf{0.0025} 	& \textbf{0.0072} 	& -0.1290  & &  \textbf{0.0632} & \textbf{0.0451} & \textbf{0.0542}  & \textbf{0.0214}\\
   $D_{distance}$     & \textbf{0.0416} & -0.0025 & \textbf{0.0045}   & -0.1290  & &  \textbf{0.0054} & -0.0110 & \textbf{0.0286} & \textbf{0.0000}\\
  \hline\hline
  \end{tabular}}
\label{EmpiricalTestResults}
\end{minipage}
\end{center}
\end{table}

We calculate the excess returns of our strategy, measured as the difference between the average return of each individual stock in the optimal portfolio strategy and random strategy. Random strategy comprises random portfolios, defined as a randomly selected portfolio containing $10\%$ of the total stocks, under different combinations of market conditions. Since the optimal strategy changes as we use different parameters to select portfolios and different criteria to identify market conditions, the excess returns which are shown in table~\ref{EmpiricalTestResults} differ correspondingly. It can be seen from the table that, in most cases higher profits can be obtained by our strategies compared with the random strategy. Specifically, $65\%$ returns of our strategies are larger than those of the random strategy in the Shanghai A-Share market, and the proportion is $70\%$ in the Shenzhen A-Share market. Furthermore, when using $K$ as the parameter to select central or peripheral portfolios, returns of our strategies are always higher than those of the random strategy. While the portfolios selected by $C$ rarely outperform random portfolios. Since our strategy based on "AND" criterion in the Shanghai A-Share market is rarely used, the excess returns under which are mostly negative. The most profitable strategy for the Shanghai A-share market uses $D_{distance}$ as the parameter to select portfolios and identifies market conditions based on trading day criterion, which has an excess return 0.0416. For the Shenzhen A-Share market, the most profitable strategy uses $D_{correlation}$ as the parameter to select portfolios and identifies market conditions based on trading day criterion, which has an excess return 0.0632.

\begin{figure}
\begin{center}
\begin{minipage}{15cm}
\resizebox*{15cm}{!}{\includegraphics{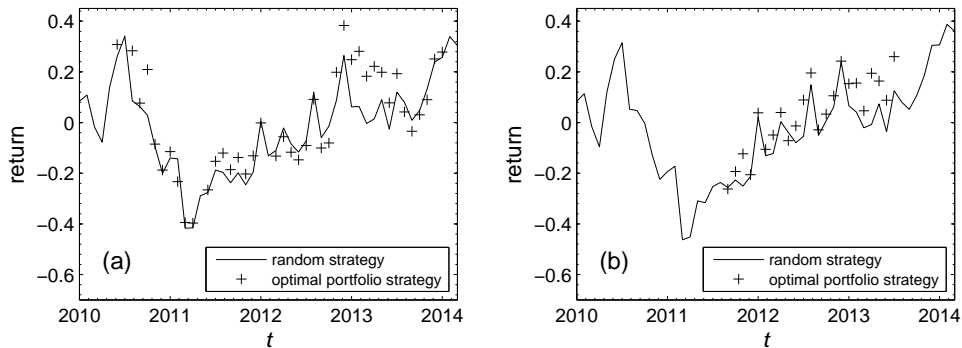}}
\end{minipage}
\end{center}
\caption{\label{Empirical Results;graph1} Average returns of the most profitable strategy (cross) and random strategy (black solid line) for the Shanghai A-Share market (a) and the Shenzhen A-Share market (b). This strategy is described in table~\ref{Empirical Results;strategy1}. }
\end{figure}

Table~\ref{Empirical Results;strategy1} lists the optimal portfolios of the most profitable strategy under different combinations of market conditions for both markets. One can see that central portfolios are chosen as the optimal portfolios under market conditions of UD, SS and DU, and peripheral portfolios are chosen under market conditions of SD for the Shanghai A-share market. For the Shenzhen A-Share market, central portfolios are chosen as the optimal portfolios under market conditions of SU and SS. The average returns of each individual stock in every investment horizon of the most profitable strategy and random strategy are plotted in figure~\ref{Empirical Results;graph1}, where the crosses are the average returns gained by the most profitable strategy in each investment horizon and the black solid line shows the average returns gained by random strategy. In 65.85\% of the investment horizons, the  average returns gained by the most profitable strategy are larger than those gained by the random strategy in the Shanghai A-Share market, and the proportion is $91.30\%$ in the Shenzhen A-Share market.

\begin{table}
\begin{center}
\begin{minipage}[c]{1\linewidth}
\tbl{ Specific optimal portfolios under particular combinations of market conditions comprise the strategy which has the highest return. The most profitable strategy for the Shanghai A-Share market uses $D_{distance}$ to select portfolios and identifies market conditions based on trading day criterion, and the most profitable strategy for the Shenzhen A-Share market uses $D_{correlation}$ to select portfolios and identifies market conditions based on trading day criterion. }
{\begin{tabular}{cccccccccc}
  \hline\hline
  & \multicolumn{4}{c}{Shanghai A-Share market} & & \multicolumn{2}{c}{Shenzhen A-Share market}\\
  \cline{2-5} \cline{7-8}
 Market condition  &  UD   & SS  & SD & DU   & & SU & SS \\
 \hline
 Optimal portfolio  &  central  & central & peripheral  & central & & central & central \\
  \hline\hline
  \end{tabular}}
\label{Empirical Results;strategy1}
\end{minipage}
\end{center}
\end{table}

\begin{figure}
\begin{center}
\begin{minipage}{15cm}
\resizebox*{15cm}{!}{\includegraphics{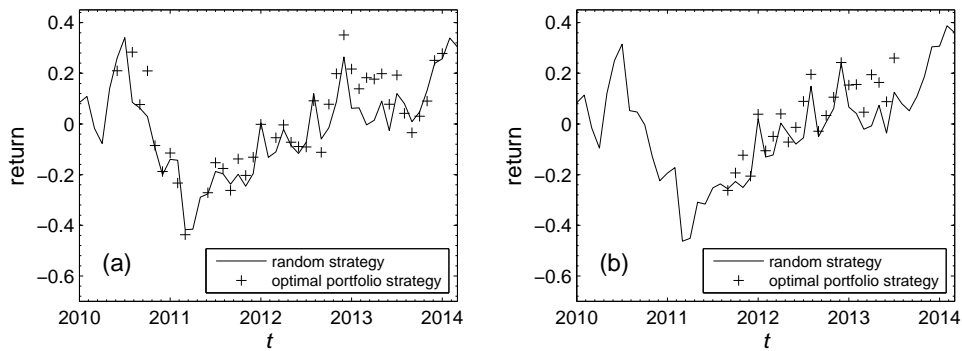}}
\end{minipage}
\end{center}
\caption{\label{Empirical Results;graph2} Average returns of the strategy (cross) that has the largest probability of gaining more profits than the random strategy (black solid line) in the Shanghai A-Share market (a) and the Shenzhen A-Share market (b). This strategy is described in table~\ref{Empirical Results;strategy2}. }
\end{figure}

We further find out the strategy that has the largest probability of gaining more profits than the random strategy, and in other words it outperforms random strategy in most of investment horizons. This strategy uses $D_{correlation}$ as the parameter to select portfolios and identifies market conditions based on trading day criterion for both two markets. The optimal portfolios under different combinations of market conditions are listed in table~\ref{Empirical Results;strategy2}. One can see that central portfolios are chosen as the optimal portfolios under market conditions of UD, SS and DU for the Shanghai A-share market. For the Shenzhen A-Share market, this strategy is also the most profitable strategy with central portfolios chosen as the optimal portfolios under market conditions of SU and SS. Similar to the above, the average returns of each individual stock in every investment horizon of this strategy and random strategy are plotted in figure~\ref{Empirical Results;graph2}. In $70\%$ of the investment horizons, the average returns gained by our strategy are larger than the random strategy in the Shanghai A-Share market, and the proportion is $91.30\%$ in the Shenzhen A-Share market.

\begin{table}
\begin{center}
\begin{minipage}[c]{1\linewidth}
\tbl{ Specific optimal portfolios under particular combinations of market conditions comprise the strategy that has the largest probability of gaining more profits than the random strategy. This strategy uses $D_{correlation}$ as the parameter to select portfolios and identifies market conditions based on trading day criterion for both markets. }
{\begin{tabular}{cccccccccc}
  \hline\hline
    & \multicolumn{3}{c}{Shanghai A-Share market} & & \multicolumn{2}{c}{Shenzhen A-Share market}\\
  \cline{2-4} \cline{6-7}
 Market condition  &  UD       & SS     & DU      & &  SU  & SS \\
 \hline
 Optimal portfolio  &  central  & central & central & & central & central\\
  \hline\hline
  \end{tabular}}
\label{Empirical Results;strategy2}
\end{minipage}
\end{center}
\end{table}

\section{Summary}

In this paper, we propose a new dynamic portfolio strategy based on the time-varying structures of MST networks for the Shanghai and Shenzhen A-Share markets. The strategy first selects central and peripheral portfolios in the selection horizon using five topological parameters and uses the selected portfolios for investment in the investment horizon. Nine combinations of market conditions haven been considered when comparing the excess returns of central and peripheral portfolios, which are identified by the ratio of the number of trading days with rising index to the total number of trading days, or the ratio of the sum of the amplitudes of the trading days with rising index to the sum of the amplitudes of the total trading days. By picking out the portfolios with larger excess returns under different combinations of market conditions, the optimal portfolios under specific market conditions have been found out: I. If the market is likely to have a drawup trend in the following investment horizon, central portfolios should be the best choice, while the peripheral portfolios usually perform worse for excessive diversification. II. If the market will be in a relatively stable state in the investment horizon, central portfolios are preferred unless the market just goes through a drawdown trend in the selection horizon. III. If the market is likely to have a drawdown trend in the investment horizon and the market is stable in the selection horizon, the peripheral portfolios should be chosen to reduce risks.

Empirical tests have also been carried out and verified the efficiency of our optimal portfolio strategy. We have used the data from 2000 to 2010 to select the optimal portfolio under each specific combination of market conditions through a training process. The selected optimal strategy have been selected according to the current market condition to make investment based on the data from 2010 to 2014. By calculating the excess returns of the optimal portfolio strategies, our strategies have been found to outperform the random strategy in most cases. Among all possible optimal portfolio strategies based on different parameters to select portfolios and different criteria to identify market conditions, $65\%$ of our optimal portfolio strategies outperform the random strategy for the Shanghai A-Share market and the proportion is $70\%$ for the Shenzhen A-Share market. Using degree $K$ as the parameter to select central or peripheral portfolios, returns of our strategies have been always higher than those of the random strategy. What's more, the excess returns of the most profitable strategies in the Shanghai and Shenzhen A-Share markets are 0.0416 and 0.0632 respectively. The strategy that has the largest probability of gaining more profits than random strategy, outperforms random strategy in $70\%$ of the investment horizons for the Shanghai A-Share market, and the proportion is $91.30\%$ for the Shenzhen A-Share market.

\section*{Acknowledgements}

We are grateful to Prof. Wei-Xing Zhou for helpful comments and suggestions. This work was partially supported by the National Natural Science Foundation (Nos. 10905023, 71131007, 71371165, 11175079 and 11501199), Fok Ying Tong Education Foundation (No. 132013), Ningbo Natural Science Foundation (No. 2015A610160), the Jiangxi Provincial Young Scientist Training Project (No. 20133BCB23017), and the Fundamental Research Funds for the Central Universities (2015).

\bibliographystyle{rQUF}
\bibliography{E:/Papers/Auxiliary/Bibliography}
\vspace{36pt}

\end{document}